\documentclass[aps,twocolumn,amssymb,amsfonts,amsmath,showpacs,final,a4paper,superscriptaddress]{revtex4-2}

\usepackage{float}
\usepackage[dvips,final]{graphicx}
\usepackage{dcolumn}
\usepackage{bm}
\usepackage[latin1]{inputenc}
\usepackage{textcomp}
\usepackage{amsmath}
\usepackage{color}
\usepackage{hyperref} 
\hypersetup{colorlinks,linkcolor={blue},citecolor={blue},urlcolor={blue}}
\usepackage{chemformula}

\begin{document}

\title{${\mathrm{\textit{In situ}}}$ preparation of superconducting infinite-layer nickelate thin films with atomically flat surface}

\author{Wenjie Sun}
\thanks{These authors contributed equally to this work.}
\affiliation{National Laboratory of Solid State Microstructures, Jiangsu Key Laboratory of Artificial Functional Materials, College of Engineering and Applied Sciences, Nanjing University, Nanjing 210093, P. R. China}
\affiliation{Collaborative Innovation Center of Advanced Microstructures, Nanjing University, Nanjing 210093,  P. R. China}

\author{Zhichao Wang}
\thanks{These authors contributed equally to this work.}
\affiliation{National Laboratory of Solid State Microstructures, Jiangsu Key Laboratory of Artificial Functional Materials, College of Engineering and Applied Sciences, Nanjing University, Nanjing 210093, P. R. China}
\affiliation{Collaborative Innovation Center of Advanced Microstructures, Nanjing University, Nanjing 210093,  P. R. China}
 
\author{Bo Hao}
\affiliation{National Laboratory of Solid State Microstructures, Jiangsu Key Laboratory of Artificial Functional Materials, College of Engineering and Applied Sciences, Nanjing University, Nanjing 210093, P. R. China}
\affiliation{Collaborative Innovation Center of Advanced Microstructures, Nanjing University, Nanjing 210093,  P. R. China}

\author{Shengjun Yan}
\affiliation{National Laboratory of Solid State Microstructures, Jiangsu Key Laboratory of Artificial Functional Materials, College of Engineering and Applied Sciences, Nanjing University, Nanjing 210093, P. R. China}
\affiliation{Collaborative Innovation Center of Advanced Microstructures, Nanjing University, Nanjing 210093,  P. R. China}

\author{Haoying Sun}
\affiliation{National Laboratory of Solid State Microstructures, Jiangsu Key Laboratory of Artificial Functional Materials, College of Engineering and Applied Sciences, Nanjing University, Nanjing 210093, P. R. China}
\affiliation{Collaborative Innovation Center of Advanced Microstructures, Nanjing University, Nanjing 210093,  P. R. China}

\author{Zhengbin Gu}
\affiliation{National Laboratory of Solid State Microstructures, Jiangsu Key Laboratory of Artificial Functional Materials, College of Engineering and Applied Sciences, Nanjing University, Nanjing 210093, P. R. China}
\affiliation{Collaborative Innovation Center of Advanced Microstructures, Nanjing University, Nanjing 210093,  P. R. China}

\author{Yu Deng}
\affiliation{National Laboratory of Solid State Microstructures, Jiangsu Key Laboratory of Artificial Functional Materials, College of Engineering and Applied Sciences, Nanjing University, Nanjing 210093, P. R. China}
\affiliation{Collaborative Innovation Center of Advanced Microstructures, Nanjing University, Nanjing 210093,  P. R. China}

\author{Yuefeng Nie}
\email[]{ynie@nju.edu.cn}
\affiliation{National Laboratory of Solid State Microstructures, Jiangsu Key Laboratory of Artificial Functional Materials, College of Engineering and Applied Sciences, Nanjing University, Nanjing 210093, P. R. China}
\affiliation{Collaborative Innovation Center of Advanced Microstructures, Nanjing University, Nanjing 210093,  P. R. China}


\begin{abstract}
Since their discovery, the infinite-layer nickelates have been regarded as an appealing system for gaining deeper insights into high temperature superconductivity (HTSC). However, the synthesis of superconducting samples has been proved to be challenging. Here, we develop an ultrahigh vacuum (UHV) ${\mathrm{\textit{in situ}}}$ reduction method using atomic hydrogen as reducing agent and apply it in lanthanum nickelate system. The reduction parameters, including the reduction temperature (${\mathrm{\textit{T}_{R}}}$) and hydrogen pressure (${\mathrm{\textit{P}_{H}}}$), are systematically explored. We found that the reduction window for achieving superconducting transition is quite wide, reaching nearly 80$^\circ$C in ${\mathrm{\textit{T}_{R}}}$ and 3 orders of magnitude in ${\mathrm{\textit{P}_{H}}}$ when the reduction time is set to 30 mins. And there exists an optimal ${\mathrm{\textit{P}_{H}}}$ for achieving the highest ${\mathrm{\textit{T}_{c}}}$ if both ${\mathrm{\textit{T}_{R}}}$ and reduction time are fixed. More prominently, as confirmed by atomic force microscopy and scanning transmission electron microscopy, the atomically flat surface can be preserved during the ${\mathrm{\textit{in situ}}}$ reduction process, providing advantages over the ${\mathrm{\textit{ex situ}}}$ CaH$_2$ method for surface-sensitive experiments.
\end{abstract} 

\maketitle

\section{Introduction} 

The square-planar structure consisting of four-coordinate low-valence ions acts as an appealing platform for investigating novel physical properties, such as high temperature superconductivity (HTSC) \cite{siegrist1988}, magnetic interactions \cite{tsujimoto2007}, and structural distortions \cite{kim2023}. Among them, the square-planar nickelates have long been regarded as a close anologue to cuprates and thought to be an ideal platform for providing insights into HTSC \cite{anisimov1999a,lee2004}. Apart from similar layered crystal structures with 3$\mathrm{\textit{d}^{9}}$ electron counts as that in cuprates, there are also different features in nickelates, such as the absence of long range antiferromagnetic ordering, weak hybridization between Ni 3$\mathrm{\textit{d}}$ and O 2$\mathrm{\textit{p}}$ orbitals and mutiband nature of Fermi surfaces.\cite{lu2021,jiang2020,botana2020,kitatani2020} And the mechanisms behind HTSC would be hopefully clarified by comparing the similarities and differences in these two systems. To date, superconductivity has been observed in square-planar nickelate thin films, including doped infinite-layer ${\mathrm{\textit{R}}}$NiO$_{2}$ (${\mathrm{\textit{R}}}$ for La, Pr and Nd) \cite{li2019,osada2020a,osada2021,zeng2021a} and reduced Ruddlesden$-$Popper (RP)-structured Nd$_{6}$Ni$_{5}$O$_{12}$ \cite{pan2022} at ambient pressure. Experimentally, the square-planar nickelate films are synthesized by de-intercalating the apical oxygens of the precursor perovskite phases with higher valence state through topotactic reduction process using either metal hydrides \cite{Michel1983,Hayward1999,Hayward2003,tsujimoto2007,li2019,lee2020,kim2023,li2020c,zeng2020} or aluminum metal \cite{wei2023,wei2023a} as the reducing agents. Comparing to the conventional CaH$_{2}$ reduction method, the usage of aluminum metal enables the ${\mathrm{\textit{in situ}}}$ synthesis of square-planar nickelates, which facilitates the ${\mathrm{\textit{in situ}}}$ characterizations. However, an extra alumina capping layer was unavoidably formed on the top of the film surface, which hinders the surface-sensitive electronic structural characterizations to some degree, such as angle-resolved photoemission spectroscopy (ARPES) and scanning tunneling microscopy (STM). On the other hand, the film and reducing agents are coupled to each other (heated/cooled together) in either of these methods, which naturally narrows the parameter space for the reduction process and further limits the film quality improvements \cite{lee2020,osada2023}.

Taking the advantage of its smallest atomic radius and moderate electronegativity, hydrogen can easily interact with solid systems with minimum perturbations. Indeed, hydrogen has been used to engineer band structures of SrTiO$_{3}$ \cite{yukawa2013} and SrIrO$_{3}$ \cite{sun2022b} in its ionic or atomic form. Here, we develop a new UHV-based reduction method to synthesis superconducting nickelates using atomic hydrogen [as schematics shown in Fig.~\ref{fig1}(a) and Fig. S1]. Comparing to the aforementioned CaH$_{2}$ or aluminum metal methods, the reduction temperature and atmosphere can be controlled separately using atomic hydrogen, which is beneficial for the reduction parameter optimizations. For example, the reduction time can be controlled with higher accuracy, since the film and reducing agents are no longer heated simultaneously, during which the reaction may be already going on. In this paper, the reduction parameters are systematically optimized according to the x-ray diffraction (XRD) and electrical transport measurements on reduced nickelate films, while the atomically flat surface is ascertained by atomic force microscopy (AFM) and scanning transmission electron microscopy (STEM).

\section{Results and Discussion}

\begin{figure}
\includegraphics[width=1.0\linewidth]{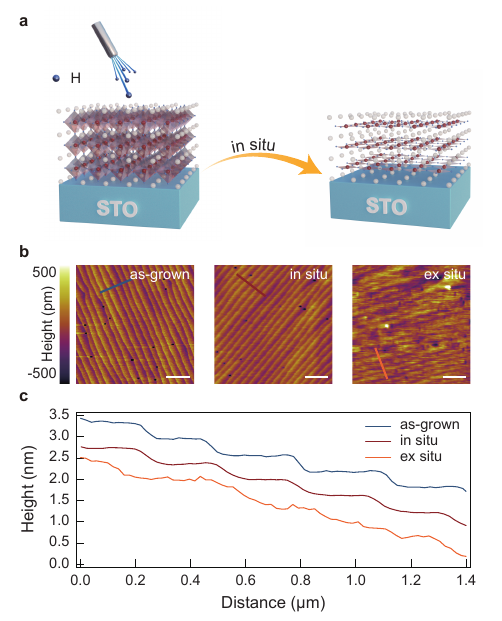}
\caption{(a) Schematic drawing shows the ${\mathrm{\textit{in situ}}}$ synthesis of superconducting nickelates using atomic hydrogen. (b) The AFM topography images of as-grown perovskite La$_{0.8}$Sr$_{0.2}$NiO$_{3}$ film and films cut from the same La$_{0.8}$Sr$_{0.2}$NiO$_{3}$ film after  ${\mathrm{\textit{in situ}}}$ and ${\mathrm{\textit{ex situ}}}$ reduction using atomic hydrogen and CaH$_{2}$, respectively. The atomically flat surface can be preserved using ${\mathrm{\textit{in situ}}}$ reduction method. The scale bars are 1 $\mu$m. (c) The corresponding line profiles along the colored solid lines in AFM images shown in (b). These curves are offset for clarity.
}
	\label{fig1}
\end{figure}

Figure~\ref{fig1}(b) shows the surface topography of nickelate films before and after the reduction process. It can be seen that the as-grown nickelate film shows step-and-terrace surface with step height of around 0.4 nm (the lattice constant of SrTiO$_{3}$) and terrace width of 0.2 $\mu$m [Fig.~\ref{fig1}(c)], nearly the same as that of the TiO$_{2}$-terminated SrTiO$_{3}$ substrate (Fig. S2). After reduction, in contrast to the roughened step-and-terrace feature for ${\mathrm{\textit{ex situ}}}$ treated film using CaH$_{2}$, the atomically flat surface can be well preserved for the ${\mathrm{\textit{in situ}}}$ treated one using atomic hydrogen. And this result proves that the atomic hydrogen causes minimum damage to the surface quality compared to the conventional CaH$_{2}$ method during the reduction process. We also note that the SrTiO$_{3}$ capping layer is thought to be necessary for preserving smooth surface of atomic-hydrogen-reduced NdNiO$_{2}$ \cite{parzyck2024}, however, our results suggest that the ${\mathrm{\textit{in situ}}}$ reduced films can retain the atomically flat surface even without any capping layers. And the reason behind such discrepancy may reside in different reduction parameters (i.e. hydrogen pressure) and Sr doping level (20 \% in our case), which could lower the oxygen removal energy \cite{goodge2023}.

\begin{figure}
\includegraphics[width=1.0\linewidth]{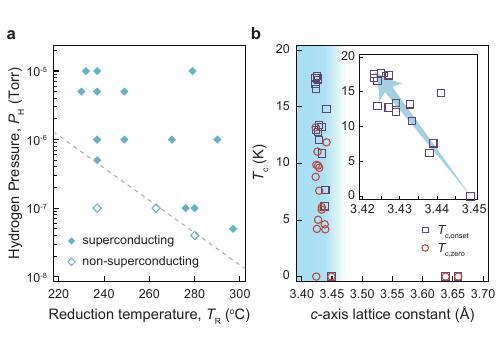}
\caption{(a) The reduction parameters of hydrogen pressure (${\mathrm{\textit{P}_{H}}}$) and reduction temperature (${\mathrm{\textit{T}_{R}}}$) used for ${\mathrm{\textit{in situ}}}$ reduction process with a fixed reduction period of 30 mins. The solid diamonds represent the successful reduction to the superconducting phase, while the hollow ones represent the unsuccessful transition. The gray dashed line is the guide to the eyes. Each symbol represents an individual reduction experiment.
(b) Superconducting transition temperature (${\mathrm{\textit{T}_{c}}}$) as a function of ${\mathrm{\textit{c}}}$-axis lattice constant. Inset is the enlarged view at lower ${\mathrm{\textit{c}}}$-axis lattice constant and only contains ${\mathrm{\textit{T}_{c,onset}}}$ for clarity. Here and for the rest of the paper, ${\mathrm{\textit{T}_{c,onset}}}$ is defined by the criterion of 90${\mathrm{\%}}$ $\rho$${\mathrm{_{n}}}$(T), where $\rho$${\mathrm{_{n}}}$(T) is determined through the linear fit to the normal state resistivity between 20 and 25 K, where ${\mathrm{\textit{T}_{c,zero}}}$ is defined by the criterion of 1${\mathrm{\%}}$ $\rho$${\mathrm{_{n}}}$(T). One can see from the enlarged view that the smaller ${\mathrm{\textit{c}}}$-axis lattice constant corresponds to higher ${\mathrm{\textit{T}_{c,onset}}}$. The blue shaded arrow is the guide to the eyes.  
}
	\label{fig2}
\end{figure}

During the ${\mathrm{\textit{in situ}}}$ reduction process, there are three main parameters that determine the reaction evolution, namely, hydrogen pressure (${\mathrm{\textit{P}_{H}}}$), reduction temperature (${\mathrm{\textit{T}_{R}}}$), and reduction time. And the reduction follows the chemical reaction:
\begin{equation}
\ch{2 H(g) + LaNiO3(s) -> H2O(g) + LaNiO2(s).}
\end{equation}
According to above chemical equation, the reaction rate can be written in Arrhenius form \cite{fu2006} as ${\mathrm{\textit{A}}}$(${\mathrm{\textit{P}_{H}}}$)$^{\alpha}$exp$(\frac{-E_{a}}{RT_{R}})$, where ${\mathrm{\textit{A}}}$ is the pre-exponential factor, $\alpha$ is the order of this reaction, $\textit{R}$ is the molar gas constant, and $E_{a}$ is the apparent activation energy. Apparently, the reaction rate is positively related to the temperature (${\mathrm{\textit{T}_{R}}}$) and reactant concentration (${\mathrm{\textit{P}_{H}}}$), while the reaction time is not that critical than the former two. In fact, the position of the film diffraction peak varies little with extending the reduction time at certain temperature before its collapse during reduction, as reported previously \cite{osada2021}. As such, we focus on the correlation between ${\mathrm{\textit{P}_{H}}}$ and ${\mathrm{\textit{T}_{R}}}$ on a series of nickelate films at a fixed reduction time of 30 mins. The corresponding XRD 2$\theta$-$\omega$ scans and transport measurements are shown in Fig. S3 and summarized in Fig~\ref{fig2}. In Fig~\ref{fig2}(a), one can see that the reduction window is quite broad, reaching about 80$^\circ$C in ${\mathrm{\textit{T}_{R}}}$ and 3 orders of magnitude in ${\mathrm{\textit{P}_{H}}}$. And the overall trends indicate a complementary relation between these two factors, that is, ${\mathrm{\textit{T}_{R}}}$ needs to be higher when ${\mathrm{\textit{P}_{H}}}$ is decreased in terms of realizing the superconducting transition, and vice versa. This result can be straightforwardly understood by the synergistic role of these two parameters in deciding the reduction evolution. It is also found that the superconducting transition is closely related to the structural parameters, specifically the ${\mathrm{\textit{c}}}$-axis lattice constant, which is extracted from the corresponding (002) diffraction peak position. As shown in Fig~\ref{fig2}(b), only films with ${\mathrm{\textit{c}}}$-axis lattice constant smaller than 3.45 $\mathring{\mathrm A}$ show superconducting transition, and ${\mathrm{\textit{T}_{c,onset}}}$ increases as ${\mathrm{\textit{c}}}$-axis lattice constant decreases. These results may indicate the flatness of NiO$_{2}$ planes plays a significant role in superconductivity, that is, flat NiO$_{2}$ planes lead to smaller ${\mathrm{\textit{c}}}$-axis lattice constant and higher ${\mathrm{\textit{T}_{c,onset}}}$, in accordance with the absence of superconductivity in bulk samples where disordered NiO$_{2}$ planes exist \cite{li2020a,Wang2020,Puphal2021,Puphal2023}. 

\begin{figure}
\includegraphics[width=1.0\linewidth]{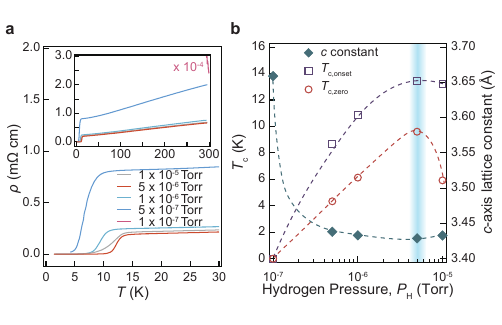}
\caption{(a) The temperature-dependent resistivity for superconducting La$_{0.8}$Sr$_{0.2}$NiO$_{2}$ films obtained under different ${\mathrm{\textit{P}_{H}}}$. ${\mathrm{\textit{T}_{R}}}$ and reduction periods are fixed at 237 $\mathrm{^{o}}$C and 30 mins, respectively. Inset shows the resistivity curves up to 300 K. Note that the resistivity curve for film reduced at ${\mathrm{\textit{P}_{H}}}$ = 1 $\times$ 10$^{-7}$ Torr is scaled by a factor of 10$^{-4}$. (b) ${\mathrm{\textit{T}_{c}}}$ and ${\mathrm{\textit{c}}}$-axis lattice constant as a function of ${\mathrm{\textit{P}_{H}}}$. Dashed lines are guides to the eyes. In this case, ${\mathrm{\textit{T}_{c}}}$ reaches its maximum at a certain ${\mathrm{\textit{P}_{H}}}$ while ${\mathrm{\textit{T}_{R}}}$ and reduction periods are fixed, while ${\mathrm{\textit{c}}}$-axis lattice constant reaches its minimum simultaneously.
}
	\label{fig3}
\end{figure}

\begin{figure}
\centering
\includegraphics[width =1.0\columnwidth]{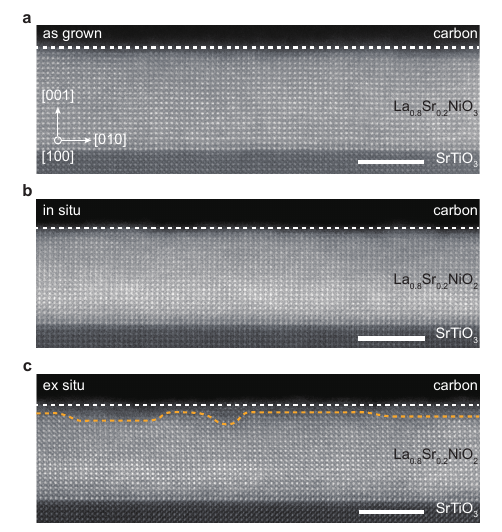}
\caption{ Cross-sectional high-angle annular dark-field scanning transmission electron microscopy (HAADF-STEM)
image of (a) perovkite La$_{0.8}$Sr$_{0.2}$NiO$_{3}$ film before reduction and infinite-layered La$_{0.8}$Sr$_{0.2}$NiO$_{2}$ films obtained using (b) ${\mathrm{\textit{in situ}}}$ and (c) ${\mathrm{\textit{ex situ}}}$ reduction method. The white dashed lines represent the nominal top surface, while the orange dashed line in (c) represents the upper boundary where well-ordered lattice can be found. It can be seen that the top surface was damaged for ${\mathrm{\textit{ex situ}}}$ reduced film, while it remains intact for ${\mathrm{\textit{in situ}}}$ reduced one. The scale bar is 5 nm. 
}
	\label{fig4}
\end{figure}

Although the superconductivity can be achieved within a quite wide reduction window, the absolute value of ${\mathrm{\textit{T}_{c}}}$ varies to some extent [Fig. S3(b) and Table S1]. Therefore, there should be an optimal correspondence between ${\mathrm{\textit{P}_{H}}}$ and ${\mathrm{\textit{T}_{R}}}$ for realizing the highest ${\mathrm{\textit{T}_{c}}}$ at a fixed reduction time. To test this hypothesis, ${\mathrm{\textit{P}_{H}}}$ was adjusted from 1$\times$10$^{-7}$ to 1$\times$10$^{-5}$ Torr at fixed ${\mathrm{\textit{T}_{R}}}$ and reduction time of 237$^\circ$C and 30 mins, respectively. As shown in Fig.~\ref{fig3}, ${\mathrm{\textit{T}_{c}}}$ changes in a non-monotonic way with raising ${\mathrm{\textit{P}_{H}}}$, and reaching its maximum at ${\mathrm{\textit{P}_{H}}}$ = 5$\times$10$^{-6}$ Torr. For ${\mathrm{\textit{P}_{H}}}$ lower than 1$\times$10$^{-7}$ Torr, the apical oxygen cannot be fully deintercalated, resulting in the La$_{0.8}$Sr$_{0.2}$NiO$_{2+\delta}$ intermediate phase with (002) diffraction peak located at around 50$^{\circ}$, as evidenced by the corresponding XRD 2$\theta$-$\omega$ scans(Fig. S4). On the other hand, the over-reduction happens when ${\mathrm{\textit{P}_{H}}}$ exceeds 5$\times$10$^{-6}$ Torr, resulting in increased normal-state resistivity and lowered ${\mathrm{\textit{T}_{c}}}$. Moreover, the ${\mathrm{\textit{c}}}$-axis lattice constant follows the opposite trend to ${\mathrm{\textit{T}_{c}}}$ as ${\mathrm{\textit{P}_{H}}}$ raises, reaching its minimum when ${\mathrm{\textit{T}_{c}}}$ is maximized. And this result again suggests that the flatness of the NiO$_{2}$ planes is potentially important for superconductivity in nickelates. Considering the synergistic relation between ${\mathrm{\textit{P}_{H}}}$ and ${\mathrm{\textit{T}_{R}}}$, the same dependence of ${\mathrm{\textit{T}_{c}}}$ should be expected for varying ${\mathrm{\textit{T}_{R}}}$ at a given ${\mathrm{\textit{P}_{H}}}$ and reduction time. As such, the systematic evolution of physical properties on the oxygen deintercalation process can be investigated through fine tuning ${\mathrm{\textit{P}_{H}}}$ (${\mathrm{\textit{T}_{R}}}$), when ${\mathrm{\textit{T}_{R}}}$ (${\mathrm{\textit{P}_{H}}}$) and reduction time are fixed.

In order to gain a deeper understanding of the structural transition during the reduction process, the high-angle annular dark-field scanning transmission electron microscopy (HAADF-STEM) was utilized as well, as shown in Fig.~\ref{fig4}. The as-grown perovskite La$_{0.8}$Sr$_{0.2}$NiO$_{3}$ shows sharp surface and interface and well ordered lattice structure. And there are no obvious traces of RP-type stacking faults for as-grown and reduced films, as evidenced by corresponding fast-Fourier-transform (FFT) images [Fig. S5]. And the overall crystallinity as well as superconducting transition (Fig. S6), which is more bulk-related, are quite close for two reduced samples. However, after ${\mathrm{\textit{ex situ}}}$ CaH$_{2}$ reduction, there exist defects region with lower contrast near the film surface, as indicated by the orange dashed line in Fig.~\ref{fig4}(a). In contrast, the surface region remains intact after the ${\mathrm{\textit{in situ}}}$ reduction process using atomic hydrogen, which is in line with the unaffected step-and-terrace feature in AFM images. As such, atomic hydrogen shows advantage in providing maximum protection in surface quality, which is beneficial for surface-sensitive measurements.

\section{Conclusions}

In conclusion, we introduce a new ${\mathrm{\textit{in situ}}}$ reduction method for synthesizing superconducting infinite-layer nickelate films using atomic hydrogen. Owing to the decoupled reduction temperature and environment, the synergistic role between them is confirmed experimentally, which also reveals a quite broad reduction window. The complete synthesis of superconducting nickelates as well as other square-planar structures could be done ${\mathrm{\textit{in vacuo}}}$ now. More importantly, the atomically flat surface could be guaranteed for ${\mathrm{\textit{in situ}}}$ reduced films, which offers great advantages for surface-sensitive structural and electronic property characterizations. This new ${\mathrm{\textit{in situ}}}$ reduction method provides an attracting pathway to understanding not only superconductivity in infinite-layer nickelates, but also novel physical properties in other square-planar structures.

\section{Materials and Methods}
$\mathrm{\textit{Film growth and reduction}}$. 
Perovskite La$_{0.8}$Sr$_{0.2}$NiO$_{3}$ thin films without capping layers were grown on TiO$_{2}$-terminated (001)-oriented SrTiO$_{3}$ substrates by reative MBE using distilled ozone as the oxidizing agent. The substrate temperature and background pressure were kept at 600$^\circ$C and 1 $\times$ 10$^{-5}$ Torr, respectively. In order to obtain the monolayer doses of La, Ni, and Sr, we have grown LaNiO$_{3}$ and SrTiO$_{3}$ using the codeposition method on SrTiO$_{3}$ substrates, and the shutter times for each element were defined from the codeposition period and optimized through RHEED oscillations \cite{li2021}.

For ${\mathrm{\textit{ex situ}}}$ reduction method using CaH$_{2}$, the perovskite films and CaH$_{2}$ powder (0.1$-$0.2 g) were together sealed in a vacuum chamber with background pressure lower than 1 $\times$ 10$^{-3}$ Torr, and then heated to 310$^\circ$C for 4 h, with warming (cooling) rate of 10$-$15$^\circ$C min$^{-1}$.

For ${\mathrm{\textit{in situ}}}$ reduction method using atomic hydrogen, the perovskite films were directly transfered into an UHV reduction chamber after growth. The base pressure of the reduction chamber is better than 1$\times$10$^{-9}$ Torr. The atomic hydrogen were generated by a home-modified e-beam source (Single Pocket Electron Beam Evaporator, SPECS) and irradiated the film surface along the normal. During the reduction process, the hydrogen gas was delivered through a leak valve into the e-beam source, and cracked into atomic hydrogen through a tungsten capillary, which is heated by electron bombardments via a tungsten filament placed in front of it (as schematics shown in Fig. S1). The accelaration voltage applied between the tungsten capillary and filament is 1.5 kV, and the filament current and emission current are 4.7 A and 70 mA, respectively. Some preliminary experiments were carried out using atomic hydrogen source designed by Fermi Instruments.

$\mathrm{\textit{Structural characterization}}$.
X-ray diffraction measurements were performed using Bruker D8 Discover diffractometer with a monochromated Cu$-$K$\alpha$ ($\lambda = 1.5418$ \AA) radiation. ${\mathrm{\textit{c}}}$-axis lattice constant was calculated from (002) peak position. Cross-sectional specimens for STEM characterizations were prepared using the standard focused ion beam (FIB) technique (Thermo Scientific Helios G4 X FIB system). Atomic-resolved HAADF-STEM images were acquired on a double spherical aberration-corrected FEI Titan G2 60-300 system operated at 300 kV.

$\mathrm{\textit{Electrical transport}}$.
The electrical transport measurements were performed either at a home-made cryostate (base temperature of 5 K) or an Oxford TeslatronPT system (base temperature of 1.5 K) using the standard van de Pauw geometry. The ohmic contacts were achieved by ultrasonic wire bonding using aluminum wires.

$\textbf{Note added}$: Atomic hydrogen treatment has been shown to be compatible with ARPES measurements in iridate system in our previous work \cite{sun2022b}, and the application in superconducting infinite-layer nickelates  will be detailed in our following work. At the mean time, during the preparation of this manuscript, we become aware that another two articles report the reduction of Nd-nickelate films using atomic hydrogen as well \cite{parzyck2023,parzyck2024}. 

\vspace{5mm}
\noindent \textbf{Acknowledgements}\\
This work was supported by the National Key R$\&$D Program of China (Grant Nos. 2021YFA1400400, and 2022YFA1402502); National Natural Science Foundation of China (Grant Nos. 11861161004, and 12381340163) and the Fundamental Research Funds for the Central Universities (Grant No. 0213$-$14380221). H. S. acknowledges the China National Postdoctoral Program for Innovative Talents (Grant No. 20230152).

\end{document}